\documentclass[%
 aps,pra,reprint,
 superscriptaddress, 
 longbibliography,noeprint,nopreprintnumbers,
 amsmath,amssymb, 
 nofootinbib,
 floatfix
]{revtex4-1}
\usepackage[utf8]{inputenc}
\usepackage[pdftex]{graphicx}
\usepackage[usenames, dvipsnames]{xcolor}

\usepackage{braket} 
\usepackage[normalem]{ulem} 
\usepackage{dcolumn}
\usepackage{bm}
\usepackage{bbold} 
\usepackage{hyperref} 
\usepackage{comment}
\usepackage{bibunits}
\defaultbibliographystyle{apsrev4-1}
\defaultbibliography{hubbard-biblio}

\newcommand{\ii}{\textrm{i}}
\newcommand{\ee}{\textrm{e}}

\newcommand{\eqr}[1]{Eq.~(\ref{#1})}
\newcommand{\fir}[1]{Fig.~\ref{#1}}

\newcommand{\tj}{{$t$-$J$\ }}
\newcommand{\tja}{{$t$-$J$-$\alpha$\ }}

\usepackage{ulem} 
\definecolor{OxfBlue}{rgb}{0, 0.333, 0.710}
\definecolor{OxfGreen}{rgb}{0, 0.710, 0.533}

\graphicspath{{./}}

\begin{document}

\def\mytitle{Anomalous spin-charge separation in a driven Hubbard system}
\title{\mytitle}

\author{Hongmin Gao}
 \email{hongmin.gao@physics.ox.ac.uk}
 \affiliation{Clarendon Laboratory, University of Oxford, Parks Road, Oxford OX1 3PU, United Kingdom}

\author{Jonathan R. Coulthard}
 \affiliation{Clarendon Laboratory, University of Oxford, Parks Road, Oxford OX1 3PU, United Kingdom}

\author{Dieter Jaksch}
 \affiliation{Clarendon Laboratory, University of Oxford, Parks Road, Oxford OX1 3PU, United Kingdom}
 \affiliation{Centre for Quantum Technologies, National University of Singapore, 3 Science Drive 2, 117543, Singapore}

\author{Jordi Mur-Petit}
 \email{jordi.murpetit@physics.ox.ac.uk}
 \affiliation{Clarendon Laboratory, University of Oxford, Parks Road, Oxford OX1 3PU, United Kingdom}
 
\date{\today}

\begin{abstract} 
Spin-charge separation (SCS) is a striking manifestation of strong correlations in low-dimensional quantum systems, whereby a fermion splits into separate spin and charge excitations that travel at different speeds.
Here, we demonstrate that periodic driving enables control over SCS in a Hubbard system near half-filling.
In one dimension, we predict analytically an exotic regime where charge travels slower than spin and can even become `frozen', in agreement with numerical calculations.
In two dimensions, the driving slows both charge and spin, and leads to complex interferences between single-particle and pair-hopping processes. 
\end{abstract}

\maketitle

\begin{bibunit}

\nocite{apsrev41Control}

\paragraph*{Introduction.}
Strongly-correlated quantum systems exhibit a plethora of interesting phenomena, such as high-$T_c$ superconductivity~\cite{Fradkin2015} or the fractional quantum Hall effect~\cite{StormerRMP99}, underpinned by a 
competition between different interactions and 
orderings of different degrees of freedom~\cite{Dagotto2005}. An example of this is the delicate interplay between magnetic and charge correlations in the ground state of lightly-doped high-$T_c$ superconductors~\cite{Gerber2015, Comin2016, daSilva2016, Julien2015} that appears very sensitive to coherent processes beyond nearest neighbours~\cite{Zheng2017, Huang2017science, Dodaro2017,  Nocera2017, Huang2018npjqm, Jiang2018, Jiang2019}.
A striking manifestation of strong fermionic correlations is spin-charge separation (SCS) \cite{Haldane1981, Auslaender2005, Jompol2009, Ma2017}, where
the elementary excitations of the system are soliton-like spin and charge (or density) excitations, of which the physical fermion appears as a composite~\cite{Putikka94, Chen1994, Weng1995, Kulic2000}.
In one-dimensional (1D) systems, SCS is predicted to occur at low energies in Luttinger liquids~\cite{Haldane1981}. Numerical simulations of the 1D Hubbard model also demonstrated SCS~\cite{Kollath2005} in a regime beyond low energy that is relevant to cold-atom implementations of the model~\cite{Recati03}. 
A typical signature of the distinct nature of spin and charge excitations in these systems is their very different propagation velocities. For instance, in the \tj model, spin excitations travel through the lattice at speed $u_s=Ja$ while the charge excitations move at speed $u_c=ta$~\cite{Schulz1991}; here $a$ is the lattice constant, $t$ the hopping energy, and $J\ll t$ the second-order exchange energy, see Eq.~\eqref{eq:tja} below.
The predictions have been confirmed in condensed-matter setups through measurements of the dispersions of the excitations \cite{Auslaender2005,Jompol2009,Ma2017}. 

In contrast to the situation in 1D, the existence of SCS in the two-dimensional (2D) Hubbard and \tj models is an open question owing partly to the lack of 2D analytical methods and partly to the limitations of current numerical methods~\cite{Putikka94, Chen1994, Weng1995, Mishchenko2001, Lauchli04, Trousselet2014}.
There is evidence that the \tj model at low fermion density is consistent with the description of a Fermi liquid \cite{Eder95} whereas at higher fillings it shows SCS with a speed of charge excitations larger than that of spin excitations~\cite{Putikka94}.

In this Letter, we demonstrate control over SCS via periodic driving of a strongly-repulsive Hubbard model near half-filling in 1D and 2D. It is known that such a system is well-described by a static \tja model \cite{Bermudez2015, Mentink2015, Bukov2016, Mendoza2017, Coulthard2017, XXXA}, where double occupancies are forbidden by the strong on-site repulsion in the underlying Hubbard system. Compared to the standard \tj model, the \tja model also includes three-site processes which play, as we show here, an important role in the dynamics. 
In 1D, we use matrix product state (MPS) methods to look at the evolution of localised spin and charge excitations of the effective \tja chain. 
We identify an exotic regime where the spin excitation speed exceeds that of the charge excitation. Interestingly, for some driving strengths before the occurrence of phase separation \cite{Emery90,Coulthard2018} we observe a
ballistic propagation of spin excitations accompanied by `freezing' of charge excitations, a phenomenon that cannot be explained by dynamic localization~\cite{Dunlap1986} or self-localization by the phase-string effect \cite{Zhu2013}. 
Moreover, the novel 'freezing' behaviour is not seen in the standard \tj chain, where the charge excitations remain mobile until phase separation occurs. 
In 2D, we perform exact diagonalization calculations on a square lattice with a spin-dependent checkerboard potential, which creates initially imbalanced density and spin profiles. After removing the potential, these imbalances oscillate in time, with different characteristic frequencies which we show can be controlled by the driving. 
These predictions can be readily tested with available experimental techniques in the field of ultracold atoms~\cite{Jotzu2014, Desbuquois2017, Messer2018, Parsons2016, Boll2016, Cheuk2016, Chiu2018, Salomon2019}, which will provide 
novel information on the interplay between density and spin degrees of freedom in strongly-interacting Hubbard systems~\cite{Zheng2017, Huang2017science, Dodaro2017,  Nocera2017, Huang2018npjqm, Jiang2018, Jiang2019},
and could assist investigations on SCS in hitherto poorly-understood regimes, such as in 2D models and high-energy excitations of 1D strongly-interacting systems.


\paragraph*{The \tja model.}
We consider a system of strongly-repulsive spin-1/2 fermions on a lattice. We describe this system with a Hubbard model $\hat{H}_{\text{Hub}} = \hat{H}_{\text{hop}}(t_0) + U \sum_{i} \hat{n}_{i\uparrow}\hat{n}_{i\downarrow}$. 
Here $\hat{H}_{\text{hop}}(t_0) = -t_0 \sum_{\langle ij \rangle \sigma} \left( \hat{c}^\dagger_{i\sigma} \hat{c}_{j\sigma} + \text{H.c.} \right)$ describes the hopping between nearest neighbour (NN) sites $\langle ij\rangle$ of a spin-$\sigma$ fermion ($\sigma=\uparrow,\downarrow$), created at site $i$ by $\hat{c}^\dagger_{i\sigma}$; 
$t_0$ is the fermion hopping amplitude between NN sites; 
$\hat{n}_{i\sigma} = \hat{c}^\dagger_{i\sigma} \hat{c}_{i\sigma}$ is the density at site $i$ of spin-$\sigma$ fermions. 
Finally, the on-site repulsion energy $U\gg t$ prevents double occupation of a single site.

We subject the system to a periodic driving of the form 
$ \hat{H}_{\text{drive}}(\tau)
 = \cos \left(\Omega \tau\right) \sum_{i} \bm{V}\cdot\bm{r}_i \hat{n}_i $, characterized by its frequency, $\Omega$, and amplitude in the $x$-$y$ lattice plane, $\bm{V}=(V_x, V_y)$;
 $\hat{n}_i = \hat{n}_{i\uparrow}+\hat{n}_{i\downarrow}$.

\begin{figure}[tb!]
\centering
 (a)~\raisebox{\dimexpr-\height+\baselineskip}{%
   \includegraphics[width=.8\linewidth]{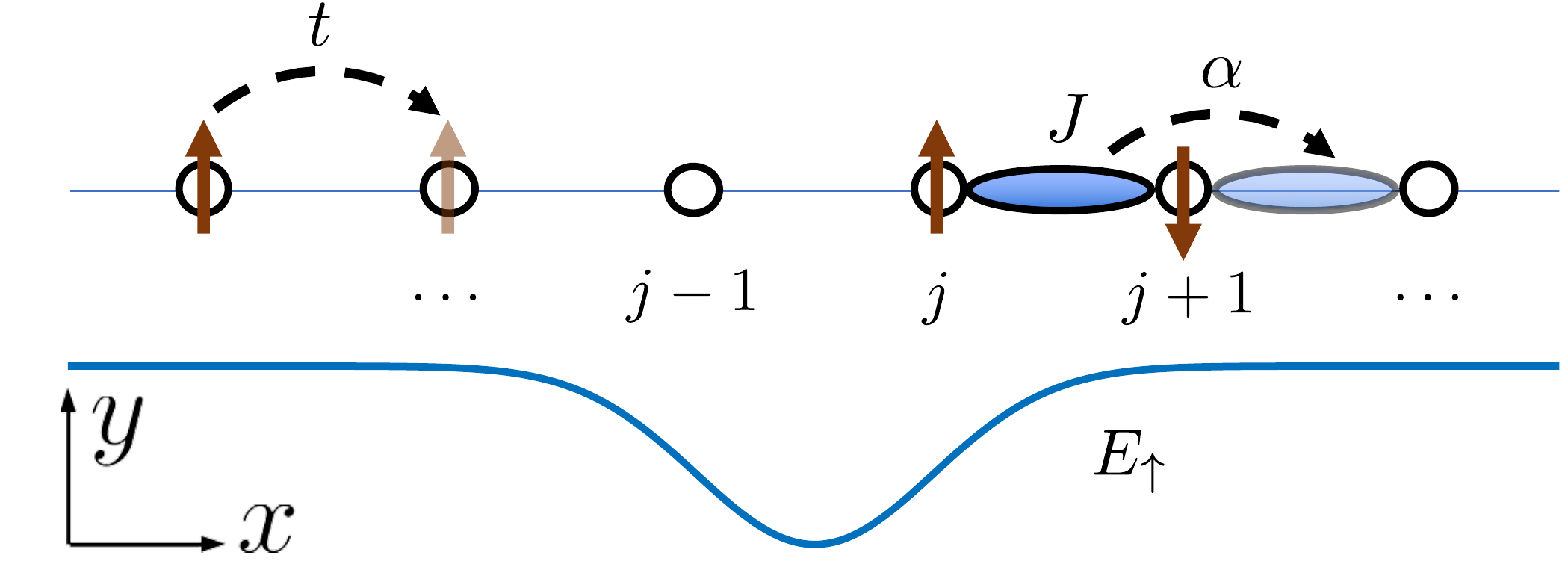}
  } 
\\
(b)~\raisebox{\dimexpr-\height+\baselineskip}{%
  \includegraphics[width=.8\linewidth]{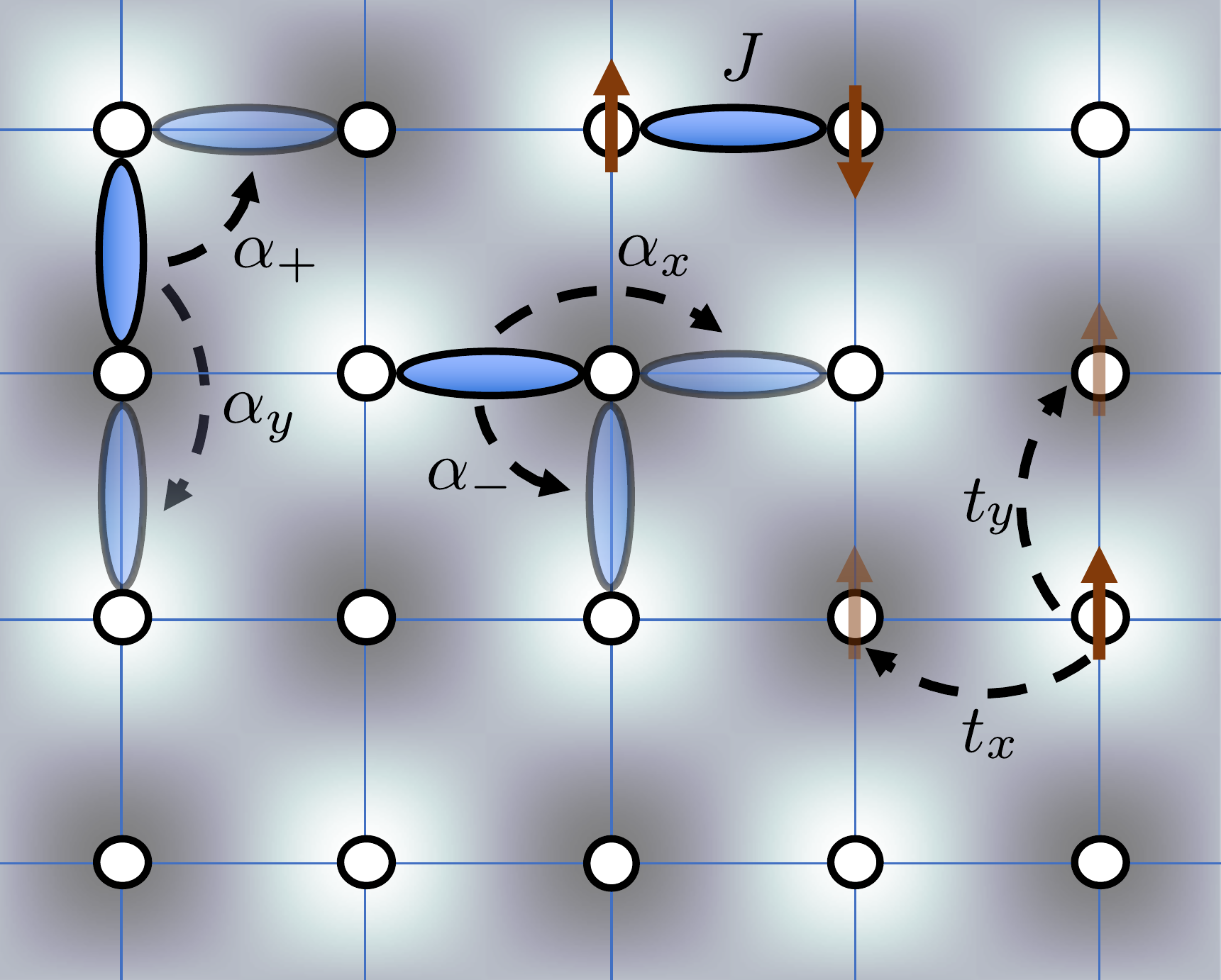}
 } 
\caption{
 (a) One-dimensional \tja chain. Spin-1/2 fermions (arrows) can hop between neighbouring sites (circles) with hopping amplitude $t$. Nearest-neighbour singlet pairs (blue ellipse) are bound by a superexchange energy $J$, and can hop from one bond to a neighbouring bond with pair-hopping amplitude $\alpha$.
 The bottom blue line illustrates the spin-dependent potential, $\hat{V}_{\text{1D}}$, felt only by the $\uparrow$ species. 
 (b) Two-dimensional \tja model. 
 Generally, the single fermion ($t_{x,y}$), and singlet-pair hopping ($\alpha_{x,y,\pm}$) amplitudes are anisotropic.
 The background shading indicates a staggered spin-dependent potential superimposed on the lattice.
 \label{fig:fig1-lattice}
}
\end{figure}

Under the condition $t_0 \ll \{ U, \Omega, |U+m\Omega|~\forall m\in\mathbb{Z} \}$, the dynamics of the driven system is described by an effective static \tja model (see \fir{fig:fig1-lattice}) \cite{Bermudez2015, Mentink2015, Bukov2016, Mendoza2017, Coulthard2017, XXXA},
\begin{equation} \label{eq:tja}
 \hat{H}_{tJ\alpha}
 = \mathcal{P}_0 \Big\lbrace 
 \hat{H}_{\text{hop}}(t) + \hat{H}_{\text{ex}}(J)
 + \hat{H}_{\text{pair}}(\{ \alpha_{ijk} \} ) \Big\rbrace \mathcal{P}_0 \,,
\end{equation}
with its parameters dependent on $\Omega$ and $\bm{V}$; see~\cite{SuppMat} for the derivations.
Here, the operator $\mathcal{P}_0 = \prod_{i} (1-\hat{n}_{i\uparrow} \hat{n}_{i\downarrow})$ projects out states with double occupancies; 
$\hat{H}_{\text{ex}}(J) = -J \sum_{\langle ij \rangle} \hat{b}^\dagger_{ij} \hat{b}_{ij}$
is the superexchange contribution, by which NN opposite spins switch their positions;
$\hat{b}^\dagger_{ij} = 
  (\hat{c}^\dagger_{i\uparrow}\hat{c}^\dagger_{j\downarrow}-\hat{c}^\dagger_{i\downarrow}\hat{c}^\dagger_{j\uparrow})/\sqrt{2}$
creates a spin-singlet pair straddling NN sites $i$ and $j$; 
$  \hat{H}_{\text{pair}}(\lbrace \alpha_{ijk} \rbrace)= - \sum^{i\ne k}_{\langle ijk \rangle} \alpha_{ijk} \hat{b}^\dagger_{ij} \hat{b}_{jk} + \text{H.c.} 
$, 
describes processes by which a singlet pair hops between nearby lattice bonds ($\langle jk \rangle \rightarrow \langle ij \rangle$), see Fig.~\ref{fig:fig1-lattice}.

\paragraph*{Anomalous SCS in one dimension.}
We consider first the case of a 1D chain with open boundary conditions, shaken with dimensionless amplitude $K=|V|/\Omega$ along its length, $L$.
Eqs.~\eqref{eq:ttilde}-\eqref{eq:alphas} below provide the
parameters of the corresponding effective \tja model: 
$t = t_0 \mathcal{J}_0(K)$,
$J = 4t_0^2 \sum_{m} \mathcal{J}_m^2(K)/(U+m\Omega)$, 
and 
$\alpha = 2t_0^2\sum_{m} \mathcal{J}_m(K) \mathcal{J}_{-m}(K)/(U+m\Omega)$.
Here ${\cal J}_m(K)$ is the $m^{\text{th}}$ order Bessel function of the first kind. 
In the limit $U \gg \Omega$, these expressions reduce to $J \approx J_0 \equiv 4t_0^2/U$ and $\alpha \approx J_0 \mathcal{J}_{0}(2K)/2$.

\begin{figure*}[t!]
\centering
\includegraphics[width = .97\linewidth]{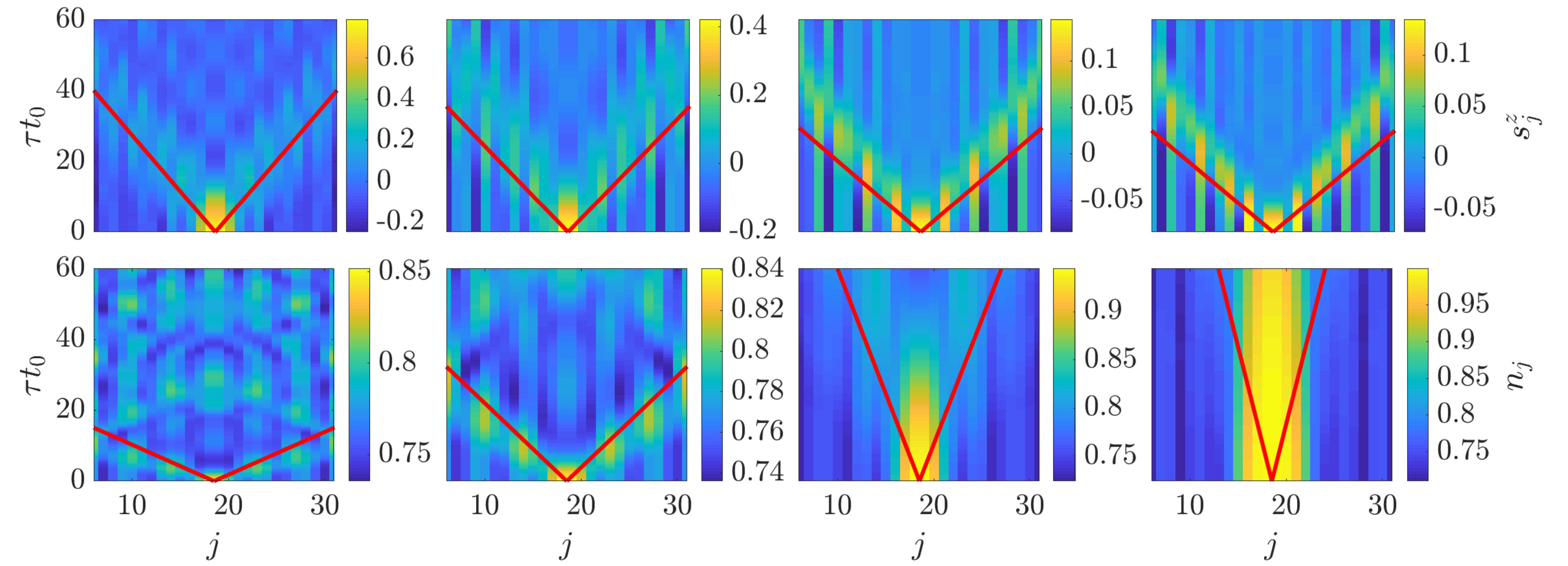}
\caption{
  Dynamics of local spin, $\langle \hat{s}^{z}_{j} \rangle$ (top row), and density, $\langle \hat{n}_{j} \rangle$ (bottom row), for the 1D \tja model as a function of position, $j$, and time, $\tau$, after the removal of the spin-dependent potential.
  From left to right, the driving strengths used are $K = 0$, $1.5$, $2.0$, and $2.1$, respectively. Note the different colorbar scales for each panel.
  The straight red lines are ballistic propagation velocity predictions from mean-field spin-charge separation (MF-SCS) theory~\cite{SuppMat}.
  Simulation parameters are: 
  $U=21t_0$ (such that  $J_0 = 0.19t_0$), $\Omega = 6t_0$, 
  $E_{\uparrow} = 0.5\times \text{max}\{|t|, J, |\alpha| \}$ and $s=2$.
  The lattice contains $L=36$ sites, of which the central 26 are shown. The total number of fermions is 28 (14 spin-$\uparrow$ + 14 spin-$\downarrow$), resulting in an average filling of $n=7/9$.
}
\label{fig:fig2-SCQuench}
\end{figure*}

To study the dynamics of spin and charge degrees of freedom in this system, following Ref.~\cite{Kollath2005} we add a weak spin-dependent potential, $  \hat{V}_{\text{1D}} = -E_{\uparrow} \sum_{j} \exp{\left[ -(j-L/2)^2/2 s^2 \right]} \hat{n}_{j, \uparrow}$, in order to create a localised spin-polarised density excitation in the centre of the lattice, see Fig.~\ref{fig:fig1-lattice}(a).
We then analyse the dynamics of the spin and density degrees of freedom upon removal of $\hat{V}_{\text{1D}}$, looking for signatures of SCS.

To start, we compute the ground state of the \tja model corresponding to the Hubbard model with given driving strength $K$, frequency $\Omega$, and spin-dependent  potential strength $E_{\uparrow}$ using the density matrix renormalisation group (DMRG) algorithm \cite{White1992, Schollwock2010}.
At time $\tau = 0$, the spin-dependent potential is switched off (while still undergoing periodic driving) and we compute the system's evolution under the effective \tja model using the time evolving block decimation (TEBD) algorithm \cite{Vidal2004, Schollwock2010}. Note that all of our simulations are performed with the \tja model, rather than the driven Hubbard model.
The validity of the \tja model as a description of the Hubbard model driven with the drive $\hat{H}_{\text{drive}}$ is established in Ref.~\cite{XXXA}.
For both DMRG and TEBD calculations, we employ the Tensor Network Theory library~\cite{AlAssam2017}. 
Our numerical results are summarised in~\fir{fig:fig2-SCQuench}, that shows the time evolution of the local spin, $s_j^z = \langle \hat{s}_j^z \rangle$, and density, $n_j = \langle \hat{n}_j \rangle$, with $\langle \hat{O} \rangle =
\langle\psi(\tau) | \hat{O} | \psi(\tau)\rangle$, $\ket{\psi(\tau)}$ being the state of the system at time $\tau$.
The leftmost column shows the undriven system, $K=0$. the  initial spin-polarised charge excitation is localised in the center of the lattice. After the spin-dependent potential is removed at time $\tau = 0$, the excitation separates into a spin excitation which propagates at a speed $u_s \approx Ja$, and a charge excitation which propagates at a higher speed $u_c \approx ta$. 
As the driving strength increases, 
Eqs.~\eqref{eq:ttilde}-\eqref{eq:alphas} predict that $t$ is suppressed while $J$ remains approximately constant.
In agreement with this prediction, our numerics shows that spin dynamics remain relatively unchanged, while the density dynamics changes drastically.
For $K \gtrsim 2$ [third and fourth columns in Fig.~\ref{fig:fig2-SCQuench}], we reach an exotic regime where spin excitations travel \textit{faster} than charge excitations. This inversion of the usual SCS scenario appears in its extreme version for $K \gtrsim 2.1$, when the charge excitation remains stationary (`freezes') in the center of the lattice despite the fact that $t \neq 0$ (in fact, $t \approx J$). 
This anomalous SCS is a robust phenomenon, as the inversion of the relative velocities of charge and spin excitations occurs for a broad range of parameters ($t_0$, $J_0$, $\Omega \ldots$).

The fact that the `freezing' happens before $t=0$ (which occurs at $K\approx 2.404$) indicates that it is a distinct phenomenon from dynamic localization~\cite{Dunlap1986}. 
We have also checked that it is not related to phase separation~\cite{Emery90,Coulthard2018} by computing the inverse compressibility, which is non-vanishing for $2.1 \lesssim K \lesssim 2.2$. 
Charge in \tj models has also been shown to localize  due to the phase-string effect~\cite{Zhu2013}, however this effect only occurs in spatial dimensions higher than 1, which excludes self-localization as an explanation for the `freezing' observed here.
Instead, we rationalize that it stems from the interplay between the direct ($t$) and spin-correlated ($\alpha$) hopping of fermions. This is supported by the fact that, if $\alpha=0$, the charge dynamics is frozen only at stronger driving, $K \gtrsim 2.3$, when phase separation occurs~\cite{Coulthard2018}.

We compare the numerical results with analytical calculations using a mean-field spin-charge separation (MF-SCS) theory based on Ref.~\cite{Feng94}.
We find that pair-hopping processes affect the charge excitation velocity, $u_c$, already at this mean-field (MF) level. Specifically, we find
\begin{equation}
    u_c
    = u_c^{t-J} 
    + 4 \alpha n \chi^2 \sin\left[ 2\pi (1-n) \right] \:,
    \label{eq: MF-SCS charge velocity}
\end{equation}
where $u_c^{t-J} = -4t\chi \sin\left[ \pi (1-n) \right]$ is the MF charge speed of the \tj model~\cite{Feng94}, $n$ is the filling fraction ($n=1$ for half-filling), and $\chi$ is the MF value of the Jordan-Wigner fermions describing neighbouring-site 
spin coherence, 
see Eq.~(S$.$20) 
in~\cite{SuppMat} for details.
At weak driving, $|\alpha| \ll t$, and $u_c$ is close to that for the standard \tj model~\cite{Feng94}. At larger drivings ($K > 1.2$), the pair-hopping ($\alpha$) terms gain in importance. As interaction terms for individual fermions, they affect the dispersions of separated spin and charge degrees of freedom [see Eq.~(S$.$26)-(S$.$27) 
in SM] such that $u_c$ is lower than in the \tj model, see Fig.~S$.$2 in~\cite{SuppMat}. These analytical predictions are in good agreement with our numerics, as shown by the solid lines in the lower panels of Fig.~\ref{fig:fig2-SCQuench}. 
We note that our MF-SCS theory predicts 'freezing' of charge excitations (see \cite{SuppMat}) even though it fails to agree with the numeric accurately on its onset (see \fir{fig:fig2-SCQuench} bottom right panel) 

Regarding the spin excitation velocity, $u_s$, 
our MF-SCS theory predicts with accuracy its value at half filling, see~\cite{SuppMat}.
For the \tj model, it is known from exact calculations that $u_s$ depends very weakly on $n$ near half-filling~\cite{Schulz1991}.
We thus follow Ref.~\cite{Feng94} and compare our MF-SCS prediction for $u_s$ at half-filling with our numerical results at $n=7/9$ in  the top panels of Fig.~\ref{fig:fig2-SCQuench}. We observe a fair agreement given the considerable assumptions of the MF treatment.
We note that, similarly to what happens in the \tj model, a fully self-consistent MF treatment overestimates the contribution of single-particle hopping to $u_s$ away from half-filling, leading to a strong $n$-dependence, see Fig.~S.1 in~\cite{SuppMat}.

\paragraph*{Anisotropic transport and SCS in two dimensions.}
We consider next the SCS scenario in a Hubbard model on a square lattice under sinusoidal time-periodic driving. 
For this case, 
the effective single-particle hopping amplitudes between NN sites $\langle ij \rangle $ separated along the $\eta=\{x,y\}$ directions are
\begin{equation}
  t_{\eta} = t_0 \mathcal{J}_0(K_{\eta}) \:,
  \label{eq:ttilde}
\end{equation}
where $K_\eta=|V_\eta|/\Omega$. 
Superexchange processes have parameters 
$J_{\eta} = 4t_0^2 \sum_{m} \mathcal{J}_m^2(K_{\eta}) / (U+m\Omega)$ for NN sites separated along $\eta=\{x,y\}$.
Finally, pair-hopping amplitudes $\alpha_{ijk}$ 
become anisotropic as well, with generally four different values, namely
\begin{align}
     \begin{array}{ll}
     \alpha_\eta = 2t_0^2 \sum_{m}
      \frac{\mathcal{J}_m(K_\eta)\mathcal{J}_{-m}(K_\eta)}
          {U+m\Omega} ,
      & \mathbf{r}_i - \mathbf{r}_k \propto \bm{\eta}=\bm{x},\bm{y} \\
     \alpha_{\pm} = 2t_0^2 \sum_{m}
      \frac{\mathcal{J}_m(K_x)\mathcal{J}_{\pm m}(K_y)}
          {U+m\Omega} ,
      & \mathbf{r}_i - \mathbf{r}_k \propto \bm{e}_{\pm}
      \end{array}
\label{eq:alphas}
\end{align}
where $\bm{e}_{\pm} = (\bm{x} \pm \bm{y})/\sqrt{2}$.

We study the system driven with dimensionless amplitudes $K_x=-K_y=K$, i.e. $\bm{V}\propto \bm{e}_-$. 
In this case, the single-particle hopping amplitudes along the $x$ and $y$ directions are suppressed equally, $t_x = t_y \equiv t = t_0{\cal J}_0(K)$ [Eq.~\eqref{eq:ttilde}], while the superexchange parameter is equal across all NN bonds, $J_x=J_x\equiv J$.
According to Eq.~\eqref{eq:alphas}, the singlet-pair hopping amplitudes are anisotropic and larger along $\bm{e}_+$:  $\alpha_x = \alpha_y = \alpha_- \neq \alpha_+$. 
For instance, in the limit $U \gg \Omega$, one has $\alpha_{-} \approx J {\cal J}_0(2K)/2$ and $\alpha_+ \approx J/2 > |\alpha_-|$. 
This anisotropy arises because under the driving a singlet-pair's potential energy changes by the same amount after hopping along $x$/$y$/$\bm{e}_-$-direction but it does not change for hopping along $\bm{e}_+$. 
To analyse the dynamics of this system with reduced finite-size and boundary effects on our results from a potentially fast-spreading localized perturbation, we impose periodic boundary conditions, and set up the initial state as the ground state of the \tja model in a weak spin-dependent potential with a checkerboard pattern,
$\hat{V}_{\text{2D}} = - E^{\text{2D}}_{\uparrow} \sum_{j_x, j_y} (-1)^{j_x + j_y} n_{j\uparrow}$, where $j=(j_x,j_y)$ labels the rows and columns of the 2D lattice and $E^{\text{2D}}_{\uparrow} $ is the strength of the potential. 
We remove this perturbing potential at time $\tau=0$, and we use exact diagonalisation for small systems to fully describe the quick growth of entanglement in the quenched system~\cite{Schollwock2010}. 
To monitor the spin and density dynamics after switching off the potential, we compute the density and spin imbalances, defined as~\cite{Trotzky2012, Schreiber2015, Landig2016, Rosson2019}
\begin{equation}
 I_{O}(\tau) = \sum_{j_x, j_y} (-1)^{j_x + j_y} \langle \hat{O}_j(\tau) \rangle ,
 \quad O=n,s \text{ .}
\end{equation}
\begin{figure} 
\centering
\includegraphics[width=.96\linewidth]{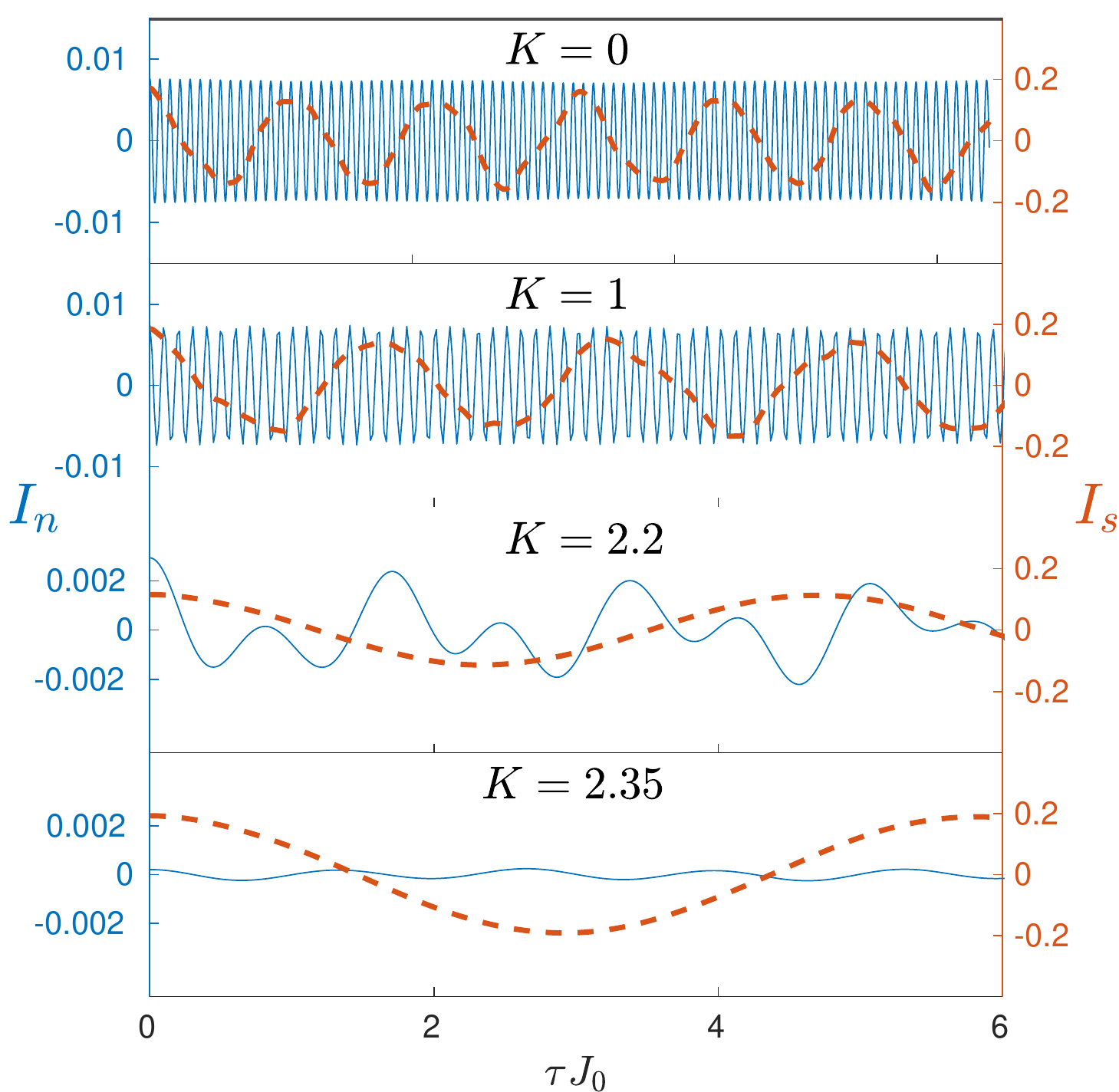}
\caption{
 Density (blue solid line, left axis) and spin (red dashed line, right axis) imbalance as a function of time for different driving strengths as indicated.
 The system simulated here is a diagonal stripe covering 12 sites of a square lattice perpendicular to the driving direction, with 5 spin-$\uparrow$ and 5 spin-$\downarrow$ fermions~\cite{SuppMat}.
 Simulation parameters are: $U=50t_0$ (such that $J_0=0.08t_0$),
 $\Omega = 14t_0$, and $E^{\text{2D}}_{\uparrow} \approx 0.05\, \max\lbrace |t|, J, |\alpha_\pm| \rbrace$.
 Note the change of left $y$-axis limits in the lower two panels.
 }
\label{fig:fig3-ImbalanceRealTime}
\end{figure}
%

We see in Fig.~\ref{fig:fig3-ImbalanceRealTime} that both $I_n$ and $I_s$ show persistent oscillations, corresponding to spin and charge excitations moving coherently between neighboring sites. Similar to the 1D case, for weak driving, $K\lesssim 1$ [top two panels in \fir{fig:fig3-ImbalanceRealTime}], the density dynamics is significantly faster than the spin dynamics.
Strong driving, $K>2$, slows down the density dynamics much more compared to spin dynamics [lower panels in \fir{fig:fig3-ImbalanceRealTime}].
In our simulations $E^{\text{2D}}_{\uparrow}$ is kept as a constant fraction of the dominant energy scale of the \tja model. 
Thus, the reduction in the amplitude of $I_n$ oscillations with increasing $K$ is due to the \tja model becoming `stiffer' to the perturbation potential.
On the other hand, the oscillation frequencies are practically unaffected by $E_{\uparrow}^{2D}$, and depend only on $K$~\cite{SuppMat}. This suggests that the changes in the spin and charge oscillation frequencies observed in 
Fig.~\ref{fig:fig3-ImbalanceRealTime} stem from the changing character of the excitations of the \tja model itself as its parameters are tuned with $K$.

While strong driving $K>2$ leads to a slowing down of density dynamics, unlike the situation in 1D we do not observe the density excitations becoming slower than the spin excitations, i.e., an inversion of the usual SCS relative speeds. In particular, it is not possible to reach the `freezing' limit in 2D.
This appears to be due to an interplay between direct and spin-correlated hoppings. 
This interplay underpins, e.g., the complex $I_n$ evolution observed for $K=2.2$ in~\fir{fig:fig3-ImbalanceRealTime}.
To understand this, we note that $\alpha_{+}(K) \approx t(K)$~\cite{XXXA} for $K\approx 2.2$, which leads to an interference between hopping events to first and second neighbours~\cite{SuppMat}. 
Numerical simulations of high-energy excitations in a lattice at low filling~\cite{XXXA}, where the dynamics is effectively described in terms of singlet pairs, also support the importance of pair-hopping terms in retaining a non-zero particle transport in 2D systems when $|t(K)| < J(K)$.
These observations are in line with recent numerical findings pointing to the relevance of next-to-nearest neighbour hopping amplitudes ($t'$) to establish the ground-state charge and spin orderings of the Hubbard model near half-filling~\cite{Zheng2017, Huang2017science, Dodaro2017,  Nocera2017, Huang2018npjqm, Jiang2018, Jiang2019}.

Finally, it is worth noting the relevance of the driving directionality: had we chosen to drive along the $x$-axis as in Ref.~\cite{Gorg2018}, $t_x$ would be renormalised but $t_y = t_0 \gg \lbrace J,|\alpha_{x,y,\pm}| \rbrace$, and single-particle hopping would dominate the dynamics, as shown in Ref.~\cite{XXXA}.

In summary, we have demonstrated that periodic driving allows one to control density (or charge) transport in low-dimensional strongly-correlated quantum systems, and to enhance the competition between direct particle transport and spin-correlated pair-hopping processes. 
In particular, we showed that in the 1D \tja model, the relative propagation speeds of the spin and charge excitations can be reversed into an exotic regime in which spin excitations travel faster than charge excitations. Moreover, we observed a regime of density `freezing' for moderately strong driving strengths, accessible by quasi-adiabatic ramping of the driving~\cite{XXXA}.
In a 2D lattice, we established that driving can lead to a severe reduction in the propagation frequencies of both spin and charge excitations, reaching a regime where coherent processes involving next-to-nearest neighbours have an enhanced impact on single-particle transport.
We expect these findings will open new routes to exploring unusual regimes of particle and spin transport, and the interplay between magnetic and superconducting correlations, in equilibrium~\cite{Corboz2014, Dolfi2015, Zheng2017, Huang2017science, Dodaro2017, Nocera2017, Huang2018npjqm, Jiang2018, Jiang2019} and out-of-equilibrium~\cite{Singla2015, Mitrano2016, Kennes2017, McIver2020, Coulthard2017, Coulthard2018, Mur-Petit2018, Buzzi2020, Budden2020, Gao2020-frank} strongly-correlated systems.

Our ideas can be implemented with existing cold-atom experimental technology~\cite{Jotzu2014,Desbuquois2017,Messer2018, Parsons2016,Boll2016, Cheuk2016,Chiu2018,Salomon2019}. This brings in the interesting possibility of tuning the effective dimensionality of the system, thus enabling one to explore in a controlled manner the role of dimensionality and anisotropy in charge and spin transport in Hubbard systems.

We would like to thank T.~Esslinger, F.~G\"org, M.~Messer, and R.\ A.\ Molina for useful discussions.
This work has been supported by EPSRC grants No.\ 
EP/P01058X/1, 
EP/P009565/1, 
and EP/K038311/1, 
the Networked Quantum Information Technologies Hub (NQIT) of the UK National Quantum Technology Programme (EP/M013243/1), 
and by the European Research Council under the European Union’s Seventh Framework Programme (FP7/2007-2013)/ERC Grant Agreement No.\ 319286 (Q-MAC). 
We acknowledge the use of the University of Oxford Advanced Research Computing (ARC) facility in carrying out this work http://dx.doi.org/10.5281/zenodo.22558. DJ partially carried out this work while visiting the Institute for Mathematical Sciences, National University of Singapore in 2019. 
HC and JRC contributed equally to this work.

\putbib
\end{bibunit}

\clearpage

\begin{bibunit}
\begin{widetext}
\begin{center}
\textbf{\large Supplemental Material: \mytitle}

H. Gao, J.R. Coulthard, J. Mur-Petit, D. Jaksch

\end{center}
\end{widetext}

\setcounter{equation}{0}
\setcounter{figure}{0}
\setcounter{section}{0}
\setcounter{table}{0}
\setcounter{page}{1}
\renewcommand\theequation{S.\arabic{equation}}
\renewcommand\thefigure{S.\arabic{figure}}
\renewcommand{\bibnumfmt}[1]{[S#1]}
\renewcommand{\citenumfont}[1]{S#1}

\section{Derivation of the effective \lowercase{$t$}-$J$-$\alpha$ Hamiltonian using Floquet basis and perturbation theory}
\label{apxsec: FullFloquetPertubativeAnalysis}

In this section we outline how to derive the effective Hamiltonian \eqr{eq:tja} governing the stroboscopic dynamics of the driven Hubbard system using Floquet theory \cite{Shirley1965, Dunlap1986, Bukov2015}. For simplicity, we restrict the discussion here to the one dimensional geometry, although the method generalises straightforwardly to higher dimensions. 

The periodic driving term in $\hat{H}_{\text{drive}}(\tau)$ breaks the continuous time-translation symmetry of the Hamiltonian to a discrete translation symmetry, meaning that energy is only conserved up to integer multiples of $\Omega$. Floquet's theorem states that due to the time-periodicity of $\hat{H}$, there exists a complete set of solutions to the time dependent Schr\"odinger equation,
\begin{equation}
| \Psi_{a}(\tau) \rangle = e^{-i\epsilon_a \tau}
|\phi_a(\tau) \rangle \text{,}
\end{equation} 
such that any state can be decomposed as a superposition of these solutions
\begin{equation}
| \Psi(\tau) \rangle = \sum\limits_{a} c_{a}
e^{-i\epsilon_a \tau} |\phi_a(\tau) \rangle .
\end{equation}
Here $|\phi_a(\tau)\rangle = |\phi_a(\tau+T)\rangle $ are time-periodic Floquet states which are solutions to the eigenvalue equation
\begin{equation}
(\hat{H} - i\partial_\tau ) |\phi_a (\tau) \rangle =
\epsilon_a |\phi_a (\tau) \rangle ,
\end{equation}
with $\epsilon_a$ termed quasienergies. The quasienergy operator, $H_{Q} = (\hat{H} - i\partial_\tau )$ acts on the combined Floquet-Hilbert space $\mathcal{H}\otimes \mathcal{T}$, where $\mathcal{H}$ is the original Hilbert space and $\mathcal{T}$ is the space of square-integrable $T$-periodic functions. The scalar product in this extended space is given by
\begin{equation}
\langle\!\langle \chi | \xi \rangle\!\rangle = \frac{1}{T}
\int\limits^{T}_{0} \text{d}\tau \langle
\chi(\tau)|\xi(\tau)\rangle ,
\end{equation}
where $| \xi \rangle\!\rangle$ denotes a vector in $\mathcal{H}\otimes \mathcal{T}$, and $|\xi(\tau)\rangle$ a $T$-periodic vector in $\mathcal{H}$. Notice that by extending the Hilbert space, we go from time-dependent matrix elements to time-independent ones. By choosing an appropriate time-periodic unitary transformation $\hat{R}(\tau)$, it is possible to bring $\hat{H}_{Q}$ into block-diagonal form, with the diagonal blocks identical up to an energy shift $m \Omega$ \cite{Eckardt2017}. The diagonal block, which acts only on $\mathcal{H}$, governs the stroboscopic dynamics of the system.

The choice of Floquet basis
\begin{equation}
|a,m \rangle = |a\rangle e^{-im\Omega \tau}
\end{equation}
conveniently structures $\mathcal{H}\otimes \mathcal{T}$ into subspaces of states that contain $m$ quanta of energy $\Omega$ from the driving. When far from resonance, blocks of different $m$ are only weakly admixed by $\hat{H}_{Q}$, so that the subspace adiabatically connected to the undriven Hubbard model is the one with $m=0$. Under the condition $t\ll \Omega \ll U$, we can perturbatively block-diagonalise $\hat{H}_{Q}$ to obtain the effective time-independent Hamiltonian.

We begin by transforming the Hamiltonian into the rotating frame with respect to the driving field. This has the effect of eliminating the explicit driving term [$\hat{H}_{\text{drive}}(\tau)$ in the main text] and imprinting it as an oscillating complex phase on the hopping term, thus bounding the terms in $\hat{H}(\tau)$ which couple different “photon" sectors by $t \ll U, \Omega$. Specifically, we transform the Hamiltonian as
\begin{equation}
    \hat{H}^{R} = \ii (\partial_{\tau} \hat{R}) \hat{R}^{\dagger} + \hat{R} \hat{H} \hat{R}^{\dagger}
\end{equation}
where
\begin{equation}
    \hat{R}(\tau) = \exp\left( \ii K \sin(\Omega \tau) \sum_{j} j \hat{n}_{j} \right).
\end{equation}
Applying this transformation to the one-dimensional driven Hubbard model, $\hat{H} = \hat{H}_{\text{Hub}} + \hat{H}_{\text{drive}}(\tau)$, we obtain 
\begin{equation}
    \hat{H}^{R}(\tau) = U \sum_{j} \hat{n}_{j, \uparrow} \hat{n}_{j, \downarrow} 
    -t_0 \sum_{j, \sigma} \left( \ee^{-\ii K \sin(\Omega \tau)} \hat{c}^{\dagger}_{j, \sigma} \hat{c}_{j+1, \sigma}  \right).
\end{equation}
The quasienergy operator in this basis is then
\begin{equation}
    \hat{H}_{\rm Q} = \left[\hat{H}^{R}(\tau) - \ii \partial_{\tau} \right].
\end{equation}
We then introduce the Floquet basis
\begin{equation} \label{eq:floquetbasis}
    \ket{a,n_d,m} =  \ket{a, n_d} \ee^{-\ii m \Omega \tau}.
\end{equation}
Here $m$ is the integer “photon number", ${-\infty < m < \infty}$, labelling the number of excitations from the periodic driving field, $n_{d}$ is the number of doubly occupied sites in the state. The remaining label $a$ denotes an arbitrary choice of basis states consistent with the labels $n_d$ and $m$. 
When we expand $\hat{H}_{\rm Q}$ in the basis in \eqr{eq:floquetbasis}, we obtain 
\begin{equation} 
    \hat{H}_{Q} = \sum_{m,m'} \hat{H}_{m,m'} \otimes \ket{m} \bra{m'},
\end{equation}
where
\begin{align} \label{eq:tbfourier2}
 \hat{H}_{m,m'}
 &=
 -t_0 \sum_{j, \sigma} \Big( \mathcal{J}_{m'-m} ( \nu ) \hat{c}^{\dagger}_{j, \sigma} \hat{c}_{j+1, \sigma} \nonumber \\
 & + \mathcal{J}_{m-m'} ( \nu ) \hat{c}^{\dagger}_{j+1, \sigma} \hat{c}_{j, \sigma} \Big) 
  + (\hat{H}_{U} +  m \Omega) \delta_{m,m'} \nonumber \\
 &= -t_0 \hat{T}_{m,m'} + (\hat{H}_{U} +  m \Omega) \delta_{m,m'}
\end{align}
are blocks which act solely on $\mathcal{H}$, while $\ket{m} \bra{m'}$ acts only on $\mathcal{T}$. We note that when $t = 0$, the quasienergy operator is trivially diagonalised by states which have well-defined $n_{d}$ and $m$. We now examine the effect of adding a finite $t \ll U, \Omega,  |U+m\Omega| ~\forall m$ as a perturbation. In general, the effective Hamiltonian within a degenerate manifold of states of a Hamiltonian $\hat{H}^{(0)}$ split by a perturbing Hamiltonian $\lambda \hat{H}^{(1)}$ is 
\begin{equation} \label{eq:geneff}
    \hat{H}_{\rm eff} = E_{n} \mathcal{P}_n + \lambda \mathcal{P}_{n} \hat{H}^{(1)} \mathcal{P}_{n} + \lambda^2 \sum_{m \neq n} \frac{\mathcal{P}_{n} \hat{H}^{(1)} \mathcal{P}_{m} \hat{H}^{(1)} \mathcal{P}_{n}}{E_{n} - E_{m}}.
\end{equation}
Here $E_{n}$ is the unperturbed energy of the $n$th degenerate manifold, which in the case of the driven Hubbard model is $E_{n_d, m} = n_d U + m \Omega$. The corresponding projector onto the $n$th degenerate manifold is $\mathcal{P}_n$. Then, making the identification
\begin{equation}
    \lambda \hat{H}^{(1)} = -t_0 \sum_{m, m'} \hat{T}_{m, m'} \otimes \ket{m} \bra{m'},
\end{equation}
and plugging this into \eqr{eq:geneff}, one obtains that the effective Hamiltonian for the manifold connected to the ground state of the strongly repulsive Hubbard model is given by
\begin{eqnarray}
    H_{\rm eff} &=& -t_0 \mathcal{P}_{0} \hat{T}_{0,0} \mathcal{P}_{0} \nonumber \\
                & & -t_0^2 \sum_{n_{d} > 0} \sum_{m} \frac{\mathcal{P}_{0} \hat{T}_{0,m} \mathcal{P}_{n_{d}} \hat{T}_{m,0} \mathcal{P}_{0}}{n_d U + m \Omega},
\end{eqnarray}
which simplifies to the \tja Hamiltonian given in \eqr{eq:tja} in the main text.

We note that this treatment breaks down when close to resonance $U \approx m\Omega$, as we no longer satisfy the condition $t \ll |U + m \Omega|$ for some choices of $m$, and states of different photon numbers become strongly admixed. 


\section{Mean-field spin-charge separation theory of 1D \lowercase{$t$}-$J$-$\alpha$ model}
In this section, we present details of our mean-field spin-charge separation (MF-SCS) theory applied to the 1D \tja model. We follow the steps outlined in Ref.~\cite{Feng94}. The first step is to transform the fermionic operators under the constraint that there is no doubly-occupied site (due to strong on-site repulsion of the underlying Hubbard model) into spinless-fermion and bosonic spin operators:
\begin{align}
\mathcal{P}_0 c^{\dagger}_{i,\uparrow}\mathcal{P}_0  &= P_i a^\dagger_i S_i^+ P_i^\dagger \text{, } & \mathcal{P}_0 c^{\dagger}_{i,\downarrow}\mathcal{P}_0  &= P_i a^\dagger_i S_i^- P_i^\dagger \nonumber\\
\mathcal{P}_0 c_{i,\uparrow}\mathcal{P}_0  &= P_i a_i S_i^- P_i^\dagger \text{, } &\mathcal{P}_0  c_{i,\downarrow}\mathcal{P}_0  &= P_i a_i S_i^+ P_i^\dagger \:.
\end{align}
Here $a_i$ is a spinless fermionic annihilation operator, $S^+_i$ is a hard-core boson creation operator, and
$P_i$ is the projection operator introduced in \cite{Feng94} that maps the three-dimensional one-site basis of the \tja model, $\{ \lbrace \ket{vac},\ket{\uparrow},\ket{\downarrow} \}$, to the fermion-spin basis.
Next, we transform the boson into another spinless fermion through a Jordan-Wigner transformation:
\begin{align}
S^{+}_j &= f_j^\dagger \exp(i\pi \sum_{l<j} f^\dagger_l f_l)\\
S^{-}_j &= f_j \exp(-i\pi \sum_{l<j} f^\dagger_l f_l)\\
S^{z}_j &= f^\dagger_j f_j - \frac{1}{2} \:.
\end{align}

We proceed to decoupling the \tja Hamiltonian in the same spirit as in Ref.~\cite{Feng94} using the following auxiliary fields for the nearest $(\eta=\pm1)$ and next-nearest $(\eta=\pm2)$ neighbor coherence,
\begin{align}
\chi_{i,i+\eta} &= f^\dagger_i f_{i+\eta} \label{eq:chi}\\
\phi_{i,i+\eta} &= a^\dagger_i a_{i+\eta} \:.
\end{align}
The mean-field approximation (MFA) amounts to replacing $\chi_{i,i+\eta}$ and $\phi_{i,i+\eta}$ by their MF values $\chi$ and $\phi$ respectively. Thus to the level of MFA, the standard \tj terms give
\begin{multline}
    H_{\text{\tj}} \approx \sum\limits_{k} \epsilon^{\text{\tj}}_{c}(k) a^\dagger_k a_k + \sum\limits_{k} \epsilon^{\text{\tj}}_{s}(k) f^\dagger_k f_k \\
    + 4Nt\chi \phi + NJ\chi^2(n^2-\phi^2) \text{,}
\end{multline}
where
\begin{align}
 \epsilon^{\text{\tj}}_{c}(k) & = -4t\chi \cos(k) - \mu\\
 \epsilon^{\text{\tj}}_{s}(k) & = \left[ J(n^2 - \phi^2)(1-2\chi) - 4t\phi \right]\cos(k)
\end{align}
are charge and spin single-particle excitation spectra.

The $\alpha$ term gives
\begin{align}
 H_{\alpha}
 &= 4\alpha n\phi\chi \sum\limits_{k} f^\dagger_k f_k \cos(k) \nonumber \\
 & \qquad + 2\alpha n\chi^2\sum\limits_{k} a^\dagger_k a_k \cos(2k) - 4\alpha N n\phi\chi^2
\end{align}

Collecting the terms together, we obtain the MF-SCS spin and charge dispersions for the \tja model:
\begin{align}
 \epsilon^{\text{\tja}}_{c}(k) & = -4t\chi \cos(k) - \mu + 2\alpha n \chi^2 \cos(2k) \label{eqn: charge dispersion}\\
 \epsilon^{\text{\tja}}_{s}(k) & = \Big\lbrack J(n^2 - \phi^2)(1-2\chi) - 4t\phi \nonumber \\
 & \qquad + 4\alpha n\phi\chi \Big\rbrack \cos(k) \:. \label{eqn: spin dispersion}
\end{align}
With these, the \tja Hamiltonian can be rewritten as
\begin{align}
 H_{\text{\tja}} &= H_{\text{\tj}} + H_{\alpha} \nonumber\\
 &=
 \sum\limits_{k} \epsilon^{\text{\tja}}_{c}(k) a^\dagger_k a_k + \sum\limits_{k} \epsilon^{\text{\tja}}_{s}(k) f^\dagger_k f_k \nonumber \\
 &~~ + 4Nt\chi \phi + NJ\chi^2(n^2-\phi^2) - 4\alpha N n\phi\chi^2 
\label{eq: MF H_tja}
\end{align}

We obtain the charge and spin excitation velocities as $u_{c (s)}(k) = \partial \epsilon^{\text{\tja}}_{c (s)}/\partial k$, with the saddle-point values of the mean fields $\chi,\phi$. The result for $u_c$ is quoted in full in \eqr{eq: MF-SCS charge velocity} in the main text. 
We plot them for different filling fractions with different driving-renormalised parameters in \fir{fig:figS1-chargespin velocity two n}. We see that the MF-SCS theory agrees that the driving slows down charge excitations. It also predicts their velocities should vanish before the driving renormalises the single-fermion hopping to zero ($K\approx2.405$ as in dynamic localization~\cite{Dunlap1986}), which is consistent with our numerics.

\begin{figure}[tb]
\centering
\includegraphics[width=0.9\linewidth,angle=0]{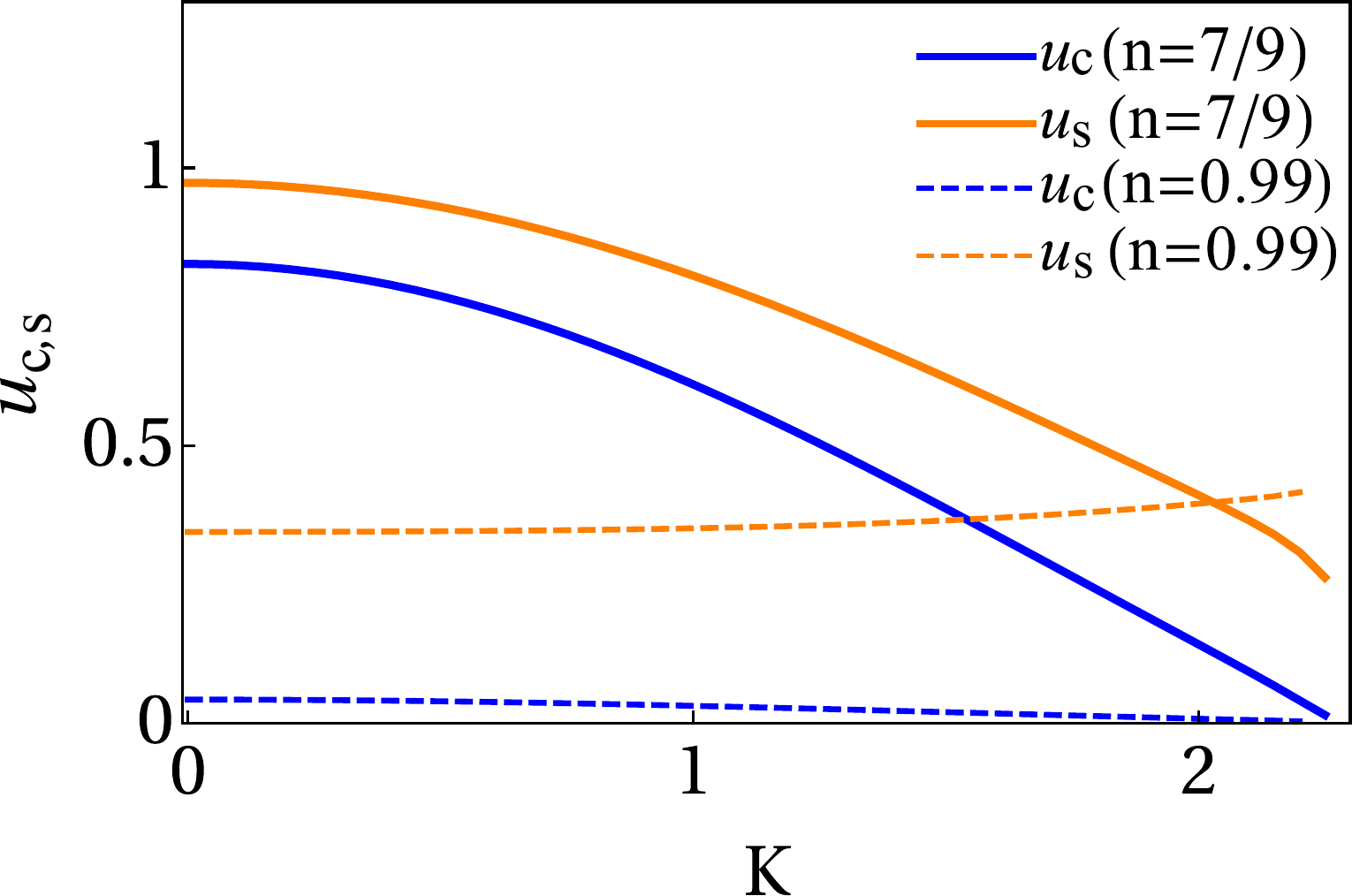}
\caption{MF-SCS theory charge and spin excitation velocities for the \tja model at two different filling fractions are plotted against the driving strength $K$. The MF-SCS theory becomes worse at predicting the spin excitation velocity when the fermion density is further away from half-filling. }
\label{fig:figS1-chargespin velocity two n}
\end{figure}

\fir{fig:figS1-chargespin velocity two n} shows that the MF-SCS theory predicts the $u_s$ varies significantly with fermion density just below half-filling, in disagreement with exact results shown in \cite{Schulz1991}. From \eqr{eqn: spin dispersion}, we see that this failure can be attributed to the MF-SCS theory over-estimating the contribution from single-fermion hopping (as a mean-field) to the spin dispersion. This contribution is small close to half-filling (it vanishes exactly at half-filling), and thus the MF-SCS results at small $K$ are fairly close to exact values for the large-$U$ Hubbard model and the Heisenberg chain \cite{Giamarchi2004}. However, MF-SCS predictions grow much greater as the doping level increases, overestimating the contribution of $t$ to $u_s$~\cite{Feng94,Schulz1991}. As $u_s$ is known to depend only weakly on fermion density close to half-filling \cite{Schulz1991}, 
we use the MF-SCS prediction for $u_s$ at half-filling for our comparison with the numerical evolution of $\langle \hat{s}_j^z\rangle$ in the 1D \tja model in \fir{fig:fig2-SCQuench} of the main text.

We compare in \fir{fig:figS2-comparisons} our MF-SCS analytical predictions for $u_c(s)$ for the \tja model with those for the standard \tj model, which are defined analogously through $u^{\text{\tj}}_{c (s)}(k) = \partial \epsilon^{\text{\tj}}_{c (s)}/\partial k$.
We observe that the pair-hopping term has visible effects on the charge velocities, especially at strong driving strength ($K>2$). In particular, each model predicts a different driving strength at which the charge excitation velocities vanishes, in qualitative agreement with our numerics. We also see in \fir{fig:figS2-comparisons}(a) and (b) that the pair-hopping term has little effect on the spin excitation velocities.

\begin{figure}[tb]
    \centering
    \includegraphics[width=.95\linewidth]{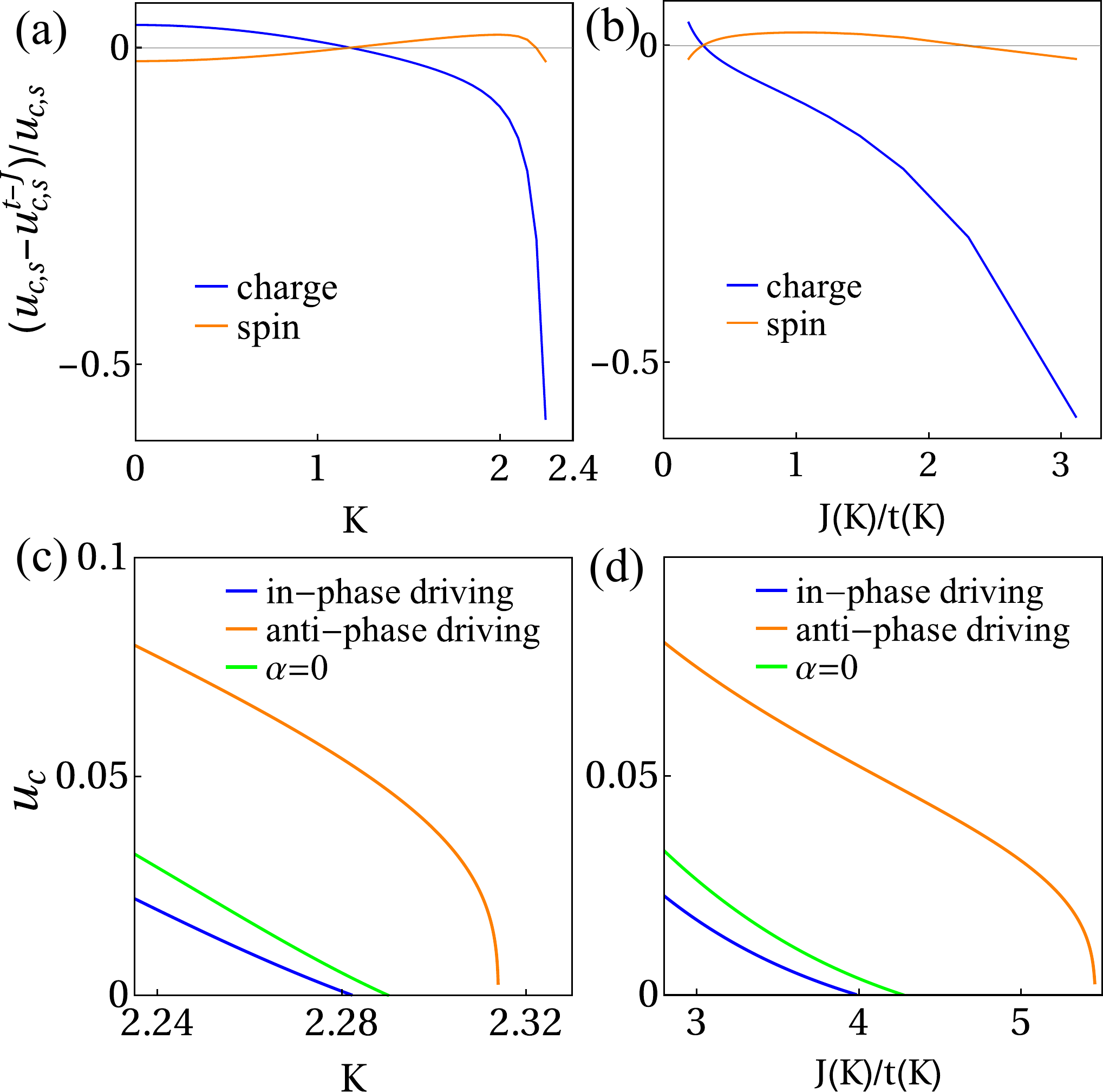}
    \caption{(a) and (b) Comparisons of the MF-SCS predictions of the spin and charge velocities between the \tj model ($u^{t-J}_{c,s}$) and \tja model ($u_{c,s}$) for the driving-renormalised parameters. (c) and (d) The charge excitation velocities according to MF-SCS theory of the \tj model ($\alpha=0$) and \tja model using in-phase (as in the main text) and anti-phase (as in \cite{Coulthard2017}) driving-renormalised parameters. (b)[(d)] shows the same quantity as (a)[(c)] but instead of plotting as a function of the driving strength $K$, we plot as a function of the ratio between the driving-renormalised parameters $J(K)$ and $t(K)$ at the same driving strengths. }
    \label{fig:figS2-comparisons}
\end{figure}

\begin{figure}[h]
    \centering
    \includegraphics[width=.95\linewidth]{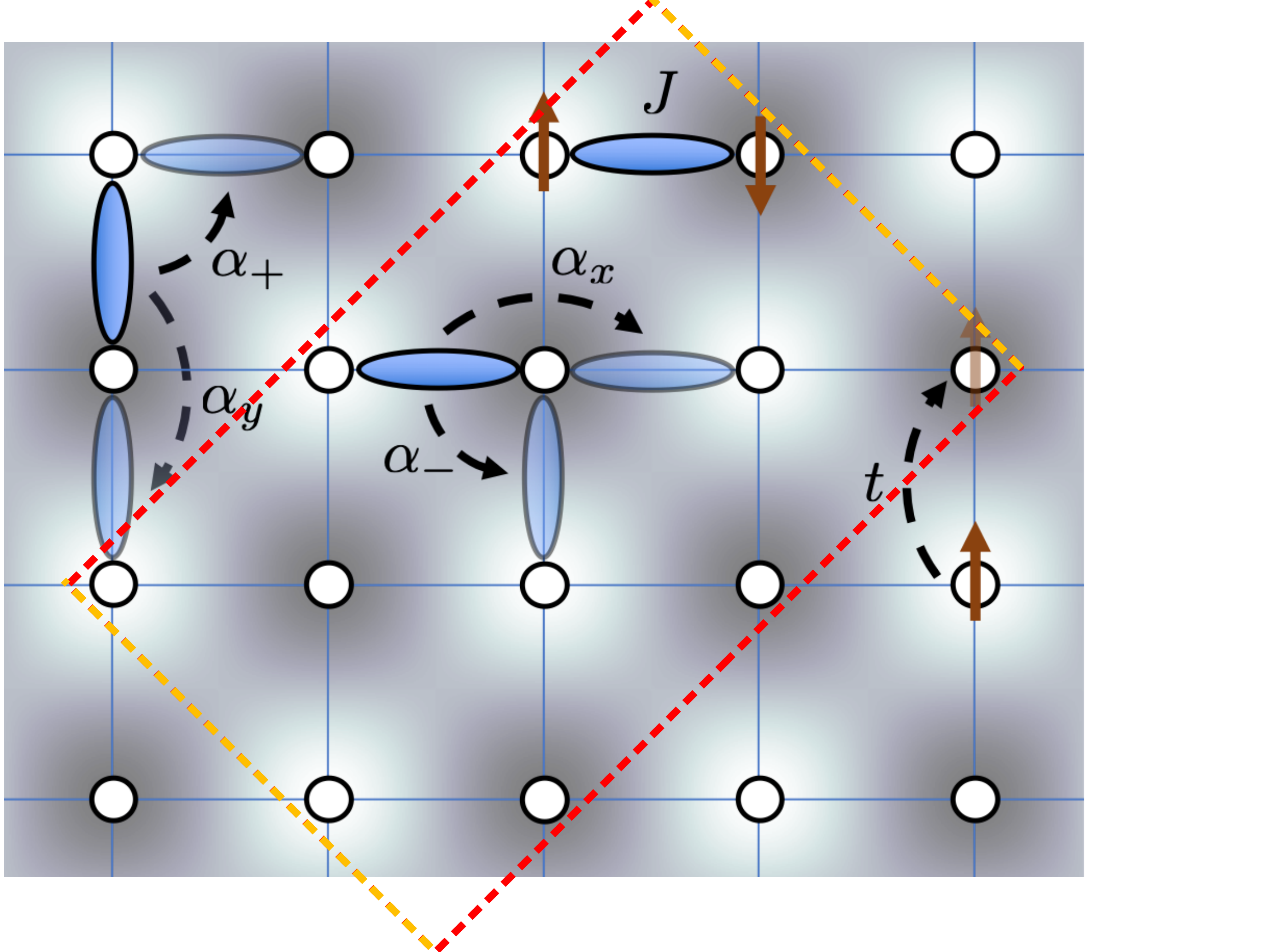}
    \caption{Lattice structure for the 2D simulations is enclosed by the dashed rectangular box. It contains 12 sites and the length is along the direction in which the pair hopping parameter is greater ($\alpha_+$ in this case). We apply periodic boundary conditions by identifying the two orange lines as well as the two red lines.}
    \label{fig:figS3-2dsimulationlattice}
\end{figure}
\section{Setup of 2D simulations}
In this section, we provide details of the setup of the 2D simulations presented in \fir{fig:fig3-ImbalanceRealTime} of the main text.

The lattice structure we simulate is a diagonal stripe of a square lattice perpendicular to the direction of periodic driving, see \fir{fig:figS3-2dsimulationlattice}. We impose periodic boundary conditions along the length and width of the diagonal stripe (identifying the two orange lines and the two red lines) such that when there is no perturbing potentials, the ground state density distribution is uniform and spin expectation value at all sites is zero. 

The initial attractive perturbing potential for spin-up particles are applied to all `even' sites (denoted by lighter shades). We perform the potential quench by first calculating the ground state with the perturbing potential. Then, we remove it at time $\tau=0$, and calculate the time evolution of the state. We ensure that the perturbing potential is weak and that the results from different driving strengths are comparable by setting the strength of the potential for each simulation to be $\approx 1/20$ of the biggest parameter amongst $t$, $J$ and $\lbrace \alpha_{x,y,\pm} \rbrace$.

\section{Interference between direct and spin-correlated hoppings in 2D systems}

\begin{figure*}[tb!]
    \centering
    \includegraphics[width=\linewidth,trim={0 0cm 0 0cm},clip]{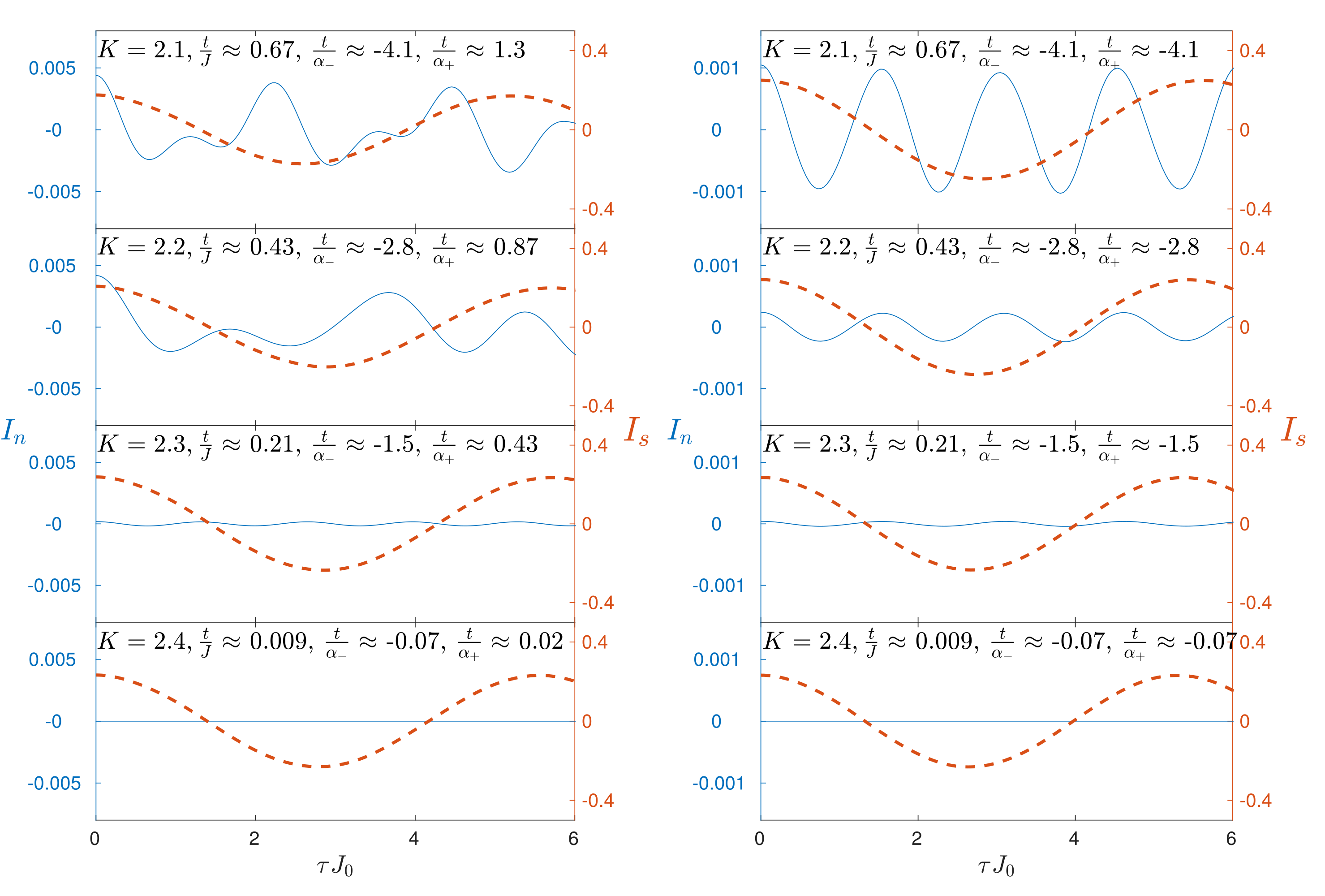}
    \caption{Time evolutions of spin and density imbalances of 2D \tja model after an initial spin-dependent perturbation on all even sites. Here we present the results with parameters calculated using four different driving strengths, $K$. $J_0 \approx 0.19 t_0$.
    In (a), parameters are calculated using $U=21 t_0$ and $\Omega=6 t_0$, driving direction is diagonal as in the main text which yields anisotropic pair-hopping parameters ($\alpha$'s), one of which is positive. While the spin imbalance does not change much with driving strength, the density imbalance reduces in amplitude and diminishes at $K\approx 2.4$ where $t \rightarrow 0$.
    In comparison, in (b), where the pair-hopping parameters are taken to be isotropic and equal to the smaller (and negative) one ($\alpha_{-}$), the density imbalance has smaller amplitude (note the different $y$-axes for density imbalances).
    Comparing the top two panels of the two figures (when $t\approx \alpha_+$), the density dynamics with anisotropic $\alpha$ is characterised by two frequencies whereas with the unphysical isotropic $\alpha$'s that are smaller in magnitude than $t$ the dynamics shows clean sinusoidal oscillations.
    }
    \label{fig:figS4-2dsimulationisotropicalpha}
\end{figure*}

In this section, we present simulations of quench experiments on 2D \tja model with  the Hubbard $U=21 t_0$ and $\Omega = 6 t_0$. 
To highlight the role the pair-hopping terms play we set the pair-hopping processes to be isotropic, i.e.\ $\alpha_{x,y,\pm}=\alpha$. 
Notice that this set of parameters is not achievable with the kind of periodic driving discussed in the main text, which produces anisotropic effective \tja Hamiltonians.

We initialise the state as the ground state of the \tja Hamiltonian with depth of the spin-dependent perturbing potential, $E^{\text{2D}}_{\uparrow} \approx 0.02 J$, before we time evolve the state without the perturbing potential. We are interested in the strong driving regime $K>2$, where $t$ is more comparable to $J$ and $\alpha$.
In \fir{fig:figS4-2dsimulationisotropicalpha}, we present numerical results of the time evolution of the spin and density imbalances, comparing models with one or both pair hopping parameters negative.

Let us compare the dynamics for $K=2.1 - 2.2$ between the model with the realisable $\alpha$'s [top two panels in the left column in Fig.~\ref{fig:figS4-2dsimulationisotropicalpha}], and an artificial model where we set $\alpha_+$ to take the (negative) value of $\alpha_-$ [right column in \fir{fig:figS4-2dsimulationisotropicalpha}].
Comparing the plots, we observe that the density dynamics with realistic, anisotropic $\alpha$'s is characterised by two frequencies.
This can be understood by noticing that, for the realistic case, for $K\approx 2.1-2.2$, we have $|\alpha_{+}(K)| \approx t(K)$, which leads to an interference effect between direct and spin-correlated hopping events.
In contrast to this, with the unphysical isotropic and smaller $\alpha$'s, the dynamics shows clean sinusoidal oscillations with a single frequency which is close to $t$.

\putbib
\end{bibunit}

\end{document}